\documentclass[%
 reprint,
 amsmath,amssymb,
 aps,
]{revtex4-2}

\usepackage{graphicx}
\usepackage{dcolumn}
\usepackage{bm}

\begin{document}

\preprint{APS/123-QED}

\title{Control of spatiotemporal localization of infrared pulses in gas-filled capillaries using weak ultraviolet pulses}

\author{Lize Han}
\author{Xiaohui Gao}
  \email{gaoxh@utexas.edu}
\affiliation{Department of Physics, Shaoxing University, Shaoxing, Zhejiang 312000, China}

\date{\today}

\begin{abstract}
Manipulation of intense pulse propagation in gas-filled capillaries is desirable for various high-field applications. Tuning the parameters of the driving laser pulse and the working gas is the conventional approach, and it provides limited capability of control. Here we demonstrate through numerical simulations a practical scheme to control the propagation of intense pulses. A weak ultraviolet pulse is launched into a capillary with a negative delay with respect to a main infrared pulse. The pulses begin to temporally overlap due to dispersion. As the main pulse self-compresses, the control pulse is strongly red-shifted due to cross-phase modulation. The frequency shifts of the two pulses mitigate pulse walk-off and allow an efficient coupling, substantially extending the effective interaction length. This interesting phenomenon may benefit applications such as high-order harmonic generation.
\end{abstract}

\maketitle
\section{Introduction}
Gas-filled capillaries and hollow-core fibers are attractive targets in intense laser-matter interaction because of the modal structure
imposed by the boundary conditions and the tunability of the dispersion and nonlinearity~\cite{Russell2014NP, Debord2019F}. The delicate interplay of Kerr effect, ionization, dispersion, and other effects gives rise to remarkably rich spatio-temporal dynamics of the laser pulse, which can be exploited for important applications such as high-order harmonic generation~\cite{Popmintchev2012S, Popmintchev2015S, Gebhardt2021LSA}, attosecond pulse generation~\cite{LiPRAPPl2022}, supercontinuum generation~\cite{BeetarJohn2021SA}, pulse self-compression~\cite{Travers2019NP}, and particle acceleration~\cite{KitagawaPRL2004}. 

Control of the propagation dynamics of the intense pulse with an intensity above $10^{14}$ W/cm$^2$ is highly desirable for these applications. For this purpose, extensive research has been carried out both in simulations and in experiments. The laser field in a cylindrical capillary can be considered as the superposition of multiple eigenmodes. When the plasma density is low, the pulse maintains its propagation in the fundamental mode. Monomode guiding of a laser pulse up to $10^{16}$\,W/cm$^2$ is experimentally demonstrated in capillaries filled with a helium gas of pressure up to 40 mbar~\cite{Dorchies1999PRL}. Ionization during the monomode propagation may assist dispersive wave generation~\cite{Chen2020OE}. At high laser intensity and gas pressure, ionization-induced refraction efficiently excites higher-order modes. Such multimode propagation has been studied both numerically and experimentally~\cite{Courtois2001PP,Chapman2010OE}. Under suitable laser and gas conditions, the multimode propagation leads to interesting phenomena such as quasi-phase-matched high-order harmonic generation~\cite{Zepf2007PRL}, spatio-temporal pulse compression~\cite{Anderson2014PRA, Gao2018OL}, intermodal four-wave mixing~\cite{Piccoli2021NP}, multidimensional solitary propagation~\cite{Safaei2020NP}, and nonlinear focused supercontinuum~\cite{Gao2022OL}. 

Adjusting the laser and gas parameters is the most straightforward way to control the propagation dynamics. For example, laser wavelength and gas pressure are optimized in high-order harmonic generation experiments~\cite{Popmintchev2012S, Popmintchev2015S}. Sophisticated manipulation is also explored such as employing pressure gradient of the gas to enhance the self-compression~\cite{Wan2021OE} and using necklace beams to generate visible light pulses~\cite{Crego2021OE}. Two-pulse scheme offers additional flexibility. It is used in many high-field applications including third harmonic generation~\cite{Hartinger2008APL}, high harmonic generation~\cite{Peng2017PRA}, supercontinuum generation~\cite{Demircan2013PRL}, and femtosecond filamentation~\cite{Bejot2015PRA}. The coupling of the two pulses can be due to ionization or Kerr effect. In particular, a main pulse can be controlled by a much weaker pulse due to cross-phase modulation~\cite{Tartara2015JOSAB, Ferraro2023JoLT} or quantum interference~\cite{Bejot2015PRA}, allowing a practical all-optical control of a powerful beam. 

In this paper, we numerically investigate the intense infrared pulse propagation in gas-filled capillaries and find that spatio-temporal confinement of the infrared pulse can be effectively controlled by a weak ultraviolet pulse. It has been known that the cross-phase modulation between the main pulse and the control pulse in a fiber leads to an efficient light by light control~\cite{Demircan2011PRL}. In Ref.~\cite{Demircan2011PRL}, a single-mode propagation with the Kerr nonlinearity is considered, and slowly varying envelope approximation is adopted.  Our study shows that this two-pulse approach is applicable for intense pulse propagation in the few-cycle regime where ionization is significant. This approach requires a low energy control pulse at an easily accessible wavelength. Thus, it could be a cost-effective way to optimize applications such as high-order harmonic generation.  

\section{Numerical Simulations}

Simulations are performed by numerically solving carrier-resolved, radially symmetric unidirectional pulse propagation equation (UPPE). Under the approximate leaky mode expansion~\cite{AndreasenPRE2013}, the electric fields in a cylindrical waveguide of radius $a$ are expressed in terms of the discrete mode component such that
$E(r,t)=\sum_{n}E_{n}(t)J_{0}(k_{\perp,n}r)$,
where $n$ is the mode index, $E_{n}(t)$ is modal component of the carrier-resolved complex electric field, $r$ is the radial position, $J_0(k_{\perp,n}r)$ is the zeroth-order Bessel function, $k_{\perp,n}$ = $u_n/a$ is the axial wave number, and $u_n$ is the $n^\mathrm{th}$ zeros of $J_0$. The approximation is valid at the first order under the condition $ka\gg1$~\cite{Cros2002PRE}, where $k$ is the wave number. 
In the spectral domain, the evolution of each modal component is given by~\cite{Couairon2011EPJST,Brown2019},

\begin{equation}
\frac{\partial E_{n}}{\partial z}=i[k_{z,n}(\omega)+i\alpha_{n}]E_{n}+\frac{i}{2k_{z,n}(\omega)}\frac{\omega^{2}}{\epsilon_{0}c^{2}}\left(\mathcal{P}_{n}+i\frac{\mathcal{J}_{n}}{\omega}\right).
\end{equation}
Here $z$ refers to the propagation distance, $k_{z,n}=\sqrt{k^2(\omega)-k_{\perp,n}^2}$ is the longitudinal wave number, $k(\omega)=\omega n(\omega)/c$, $\omega$ is angular frequency, $n(\omega)$ is the refractive index, $c$ is the speed of light in vacuum, $\epsilon_{0}$ is the vacuum permittivity, $\mathcal{P}$ is the nonlinear polarization, $\mathcal{J}=\mathcal{J}_{p}+\mathcal{J}_{a}$ is the total nonlinear polarization current, $\mathcal{J}_{p}$ is the one due to plasma generation and $\mathcal{J}_{a}$ is due to ionization absorption. The modal loss coefficient $\alpha_n$ is calculated using
\begin{equation}
\alpha_n=\frac12\left(\frac{u_n}{2\pi}\right)^2\frac{\lambda^2}{a^3}\frac{1+\epsilon_\mathrm{c}}{\sqrt{\epsilon_\mathrm{c}-1}}
\end{equation}
with $\epsilon_\mathrm{c}$ the dielectric constant of the cladding. The time evolution of the nonlinear polarization $\mathcal{P}(t)$, nonlinear currents $\mathcal{J}_{p}(t)$ and $\mathcal{J}_{a}(t)$, and the electron density $\rho_e(t)$ are calculated using the following system of equations, 
\begin{equation}
\mathcal{P}(t)=\epsilon_0\chi^{(3)}E(t)^3,
\end{equation}
\begin{equation}
\frac{\partial\mathcal{J}_p(t)}{\partial t}+\frac{\mathcal{J}_p(t)}{\tau_\mathrm{coll}}=\frac{e^2}{m_e}\rho_e(t)E(t),
\end{equation}
\begin{equation}
\mathcal{J}_a=\epsilon_0cn_0(\rho_0-\rho_e(t))\frac{W(|E(t)|)}{|E|^2}U_iE(t),
\end{equation}
\begin{equation}
 \frac{\partial\rho_e}{\partial t}=W(|E(t)|)(\rho_0-\rho_e).
 \end{equation} 
Here $\chi^{(3)}=4\epsilon_{0}cn_{2}n_{0}^{2}/3$ is the third order susceptibility, $n_2$ and $n_0$ are the nonlinear and linear refractive index, respectively,  $\tau_\mathrm{coll}$ is the collision time between electrons and neutral atoms, $e$ is the elementary charge, $m_e$ is the electron mass, $\rho_0$ is the initial number density of the gas, $W$ is the ionization rate, and $U_i$ is the ionization potential. 

The field is initialized at $z=0$ as a superposition of the main pulse field $E_m(t)$ and control pulse field $E_c(t)$. Both pulses are unchirped Gaussian pulses, which are given by
\begin{equation}
  E_m(t)=E_{m0}\exp[-2\ln2(\frac{t}{\tau_{m}})^{2}]\exp[-i(k_z-\omega_m)t],
  \end{equation} 
\begin{equation}
E_c(t)=E_{c0}\exp[-2\ln2(\frac{t-\tau_d}{\tau_{c}})^{2}]\exp[-i(k_z-\omega_c)t].
  \end{equation} 
Here the subscript $m$ and $c$ denote the quantities associated with the main pulse and the control pulse, respectively. $E_{m0}$ and $E_{c0}$ are the peak amplitudes of the laser field at the entrance of the capillary, $\tau_{m}$ and $\tau_{c}$ are the full width at half maximum (FWHM) pulse duration of the intensity envelope, $\omega_m=2\pi/\lambda_m$ and $\omega_c=2\pi/\lambda_c$ are the angular frequencies, $\lambda_m$ and $\lambda_c$ are the central wavelengths, and $\tau_d$ is the initial time delay between the two pulses. 

Unless otherwise specified, a 10-cm long capillary of 35-$\mu$m radius filled with 15-bar argon gas is used in our simulations. This length is the longest effective interaction length that we achieve. An argon gas is chosen because it is widely used in high-field applications. It also has no delayed Kerr nonlinearity, allowing a clean physics picture. The ionization rate at an instantaneous field is calculated using a modified PPT formula~\cite{Popruzhenko2008PRL} for the wavelength of the main pulse. The collision time $\tau_\mathrm{coll}$ is inversely proportional to the pressure and is taken as 190 fs at 1 bar. The nonlinear refractive index scales linearly with the pressure and a value of $1.1\times10^{-19}$\, cm$^2$/W is used for 1 bar~\cite{Zahedpour2015OL}, and $n_0$ is calculated using Sellmeier formula at the main pulse wavelength. The number of modes is 20, which is sufficient based on convergence check of the results. The main pulse is 60-$\mu$J, 30-fs, 2-$\mu$m with 30-$\mu$m $1/e^2$ beam radius, and the control pulse is a 5-$\mu$J, 16-fs, 400-nm with 15-$\mu$m $1/e^2$ beam radius. A negative delay $\tau_d=-40$\,fs is used.

\section{Results and Discussion}

\begin{figure}[htbp]
\centering
\includegraphics[width=0.45\textwidth]{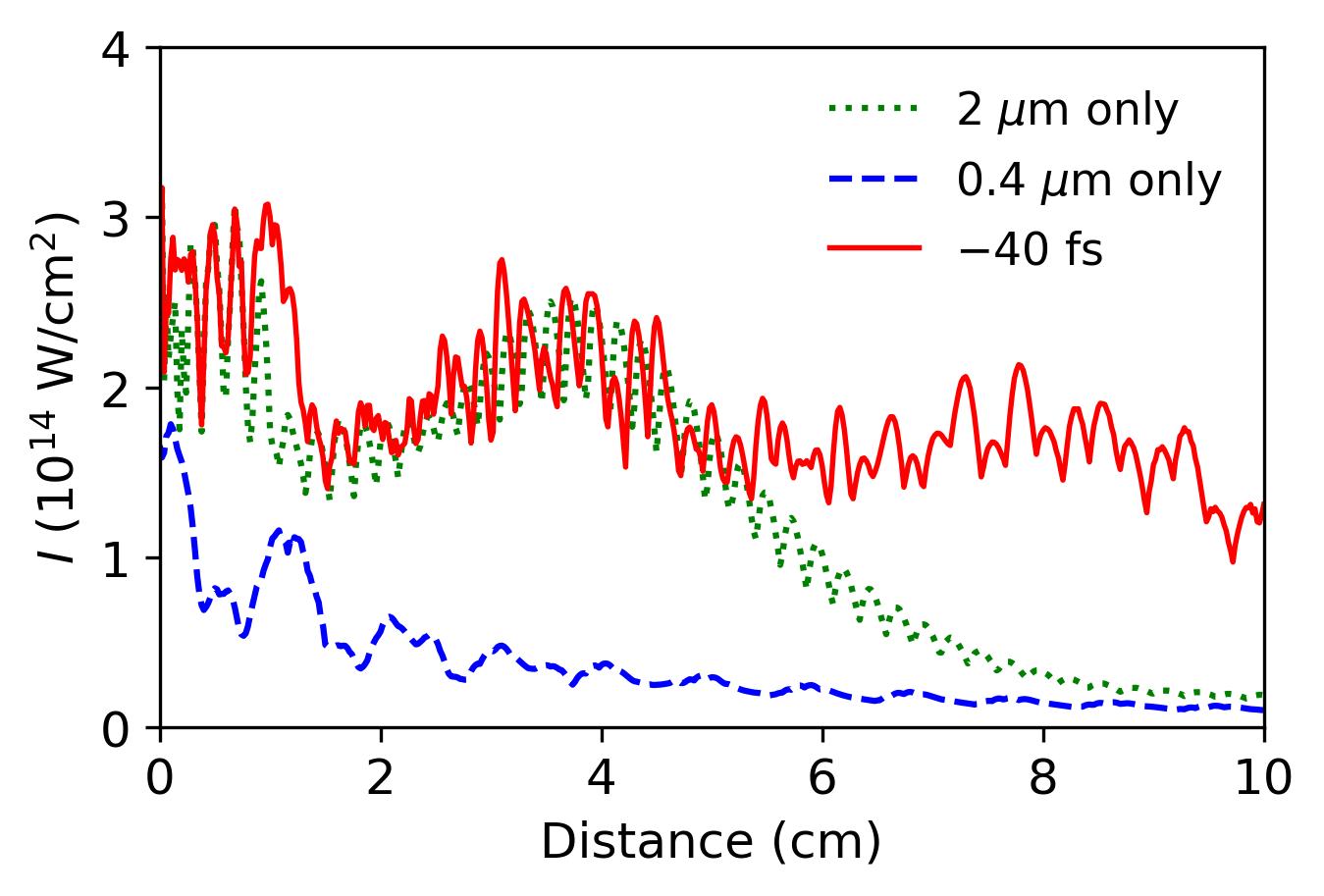}
\caption{Peak intensity versus the propagation distance for three different simulation configurations: 2-$\mu$m main pulse and 400-nm control pulse with a -40 fs delay (solid red curve), 2-$\mu$m main pulse only (dashed blue curve), and 400-nm control pulse only (dotted green curve).}
\label{fig1}
\end{figure}%

Figure~\ref{fig1} presents the peak intensity versus the distance for three different situations, demonstrating the effectiveness of controlling the main pulse by a weak pulse. The solid red curve shows the evolution of the peak intensity when the probe pulse propagates collinearly with the main pulse in the capillary with a -40 fs delay. The power of the main pulse is 1.88 GW, which is smaller than the critical power for self-focusing. The dashed blue curve and the dotted green curve show the peak intensity versus the distance when only 2-$\mu$m pulse or 400-nm pulse is present, respectively. With or without the 400-nm control pulse, the peak intensity within the first five centimeters shows similar pattern. The initial modulation is the interference of multiple modes excited during the capillary coupling. The intensity drops after the first centimeter due to the walk-off and attenuation of the higher-order modes. Owning to the self-phase modulation and anomalous group-velocity dispersion, the pulse gradually self-compresses~\cite{Balciunas2015NC}, reaching a maximum approximately near the distance of 4 cm. Further propagation brings down the intensity for the case of the main pulse only. When the control pulse is present, the intensity remains above 1.6$\times$10$^{14}$\,W/cm$^2$ for almost ten centimeter, almost doubling the effective interaction length. This striking enhancement is achieved when the energy of the control pulse is an order of magnitude lower than the main pulse.

\begin{figure}[htbp]
\centering
\includegraphics[width=0.45\textwidth]{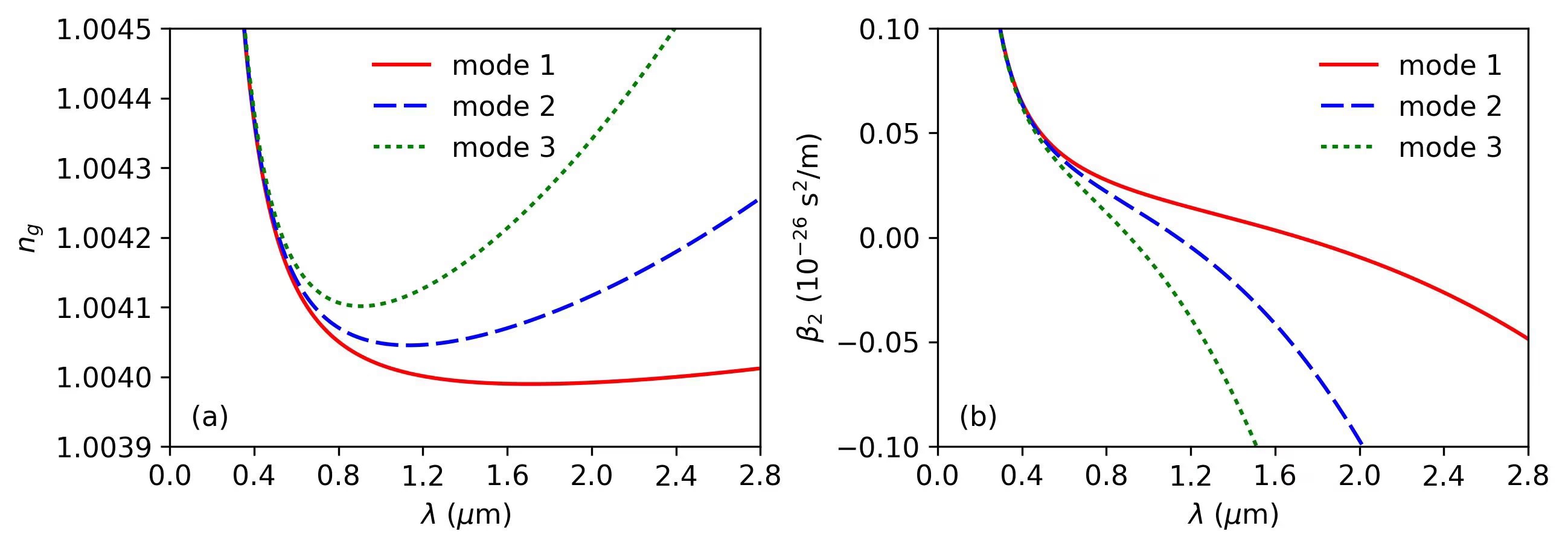}
\caption{Variation of (a) group index and (b) group-velocity dispersion with the wavelength for the lowest three modes in the capillary.}
\label{fig2}
\end{figure}%
For two-color pulse interaction, the dispersion often plays a critical role~\cite{Demircan2011PRL}. Figures~\ref{fig2}(a) and \ref{fig2}(b) show the change of group index and group-velocity dispersion with the wavelength for the lowest three modes of the capillary used in this simulation, respectively. As the wavelength increases, the group index first decreases and then rises up slowly. The wavelength for the zero group velocity dispersion for the lowest three modes are 1.71 $\mu$m, 1.13 $\mu$m, and 0.91 $\mu$m, respectively, which locates between the wavelength of the main pulse and the control pulse.  Based on these curves, the walk-off length for two 30-fs pulses of the fundamental mode at 2-$\mu$m and 400-nm is 2.4\,cm.

\begin{figure}[htbp]
\centering
\includegraphics[width=0.45\textwidth]{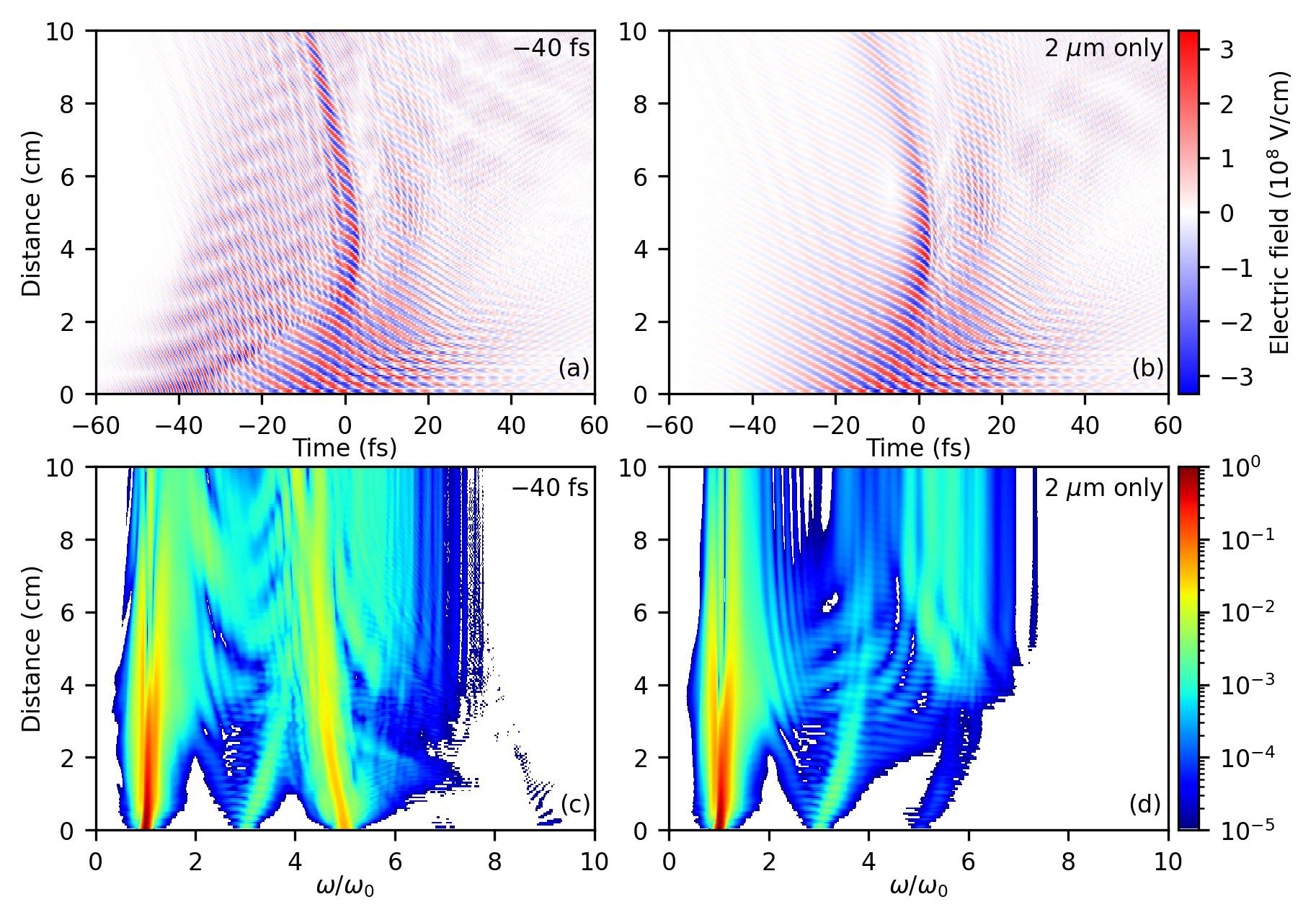}
\caption{ Comparison of the on-axis field evolution and spectral evolution with and without the control pulse. Normalized on-axis temporal profiles versus propagation distance (a) with the control pulse (b) without the control pulse. Integrated spectra versus propagation distance (c) with the control pulse (d) without the control pulse.}
\label{fig3}
\end{figure}%
To understand the role of the control pulse, the temporal evolution of the on-axis laser field without and with the control pulse are shown in Fig.~\ref{fig3}(a) and Fig.~\ref{fig3}(b), respectively. In both cases, the peak of the main pulse initially shifts to right and then shifts to left. The retarded frame moves at a reference velocity $v_g$ that is the group velocity of 2-$\mu$m. For the main pulse, the nonlinear refractive index reduces the group velocity at 2-$\mu$m. So the peak of the pulse lags behind. The self-steeping of the pulse at the trail edge gradually blue-shifts the central wavelength. After some distance, the group velocity of the blue-shifted wavelength becomes greater than that of 2-$\mu$m, allowing the pulse to catch up. The group velocity of the control pulse is smaller than $v_g$. Thus, the control pulse gets dispersed and the peak moves in the positive time. As the control pulse approaches the main pulse from the negative delay, the pulses begin to temporally overlap in the nonlinear medium and are coupled through cross-phase modulation. The effective index seen by the control pulse depends not only on its own intensity but also on the intensity of main pulse. As the intensity of the main pulse varies with the time, the resulting change of phase manifests itself as a frequency shift. This interaction is different from the two-pulse collision in the regime of optical event horizon for which the control pulse undergoes reflection. In our case, the control pulse can pass over the main pulse. Figure~\ref{fig3}(c) and Figure~\ref{fig3}(d) show the spectral evolution during the propagation. The control pulse is red-shifted because it overlaps with the leading edge of the main pulse. Without the control pulse, fifth harmonic generation also produces a weak spectral component at 400-nm. However, the harmonics are mainly generated after the peak of the pulse and is blue-shifted. The main part of the spectrum broadens up to a distance of approximately 4-cm and contracts during the further propagation due to the interplay of the self-phase modulation and group velocity dispersion, broadening the pulse in the time domain. The addition of the control pulse makes a more uniform and broader spectrum, which is essential to maintain a short pulse duration. As the main pulse is blue-shifted and the control pulse is red-shifted, both approach the zero dispersion wavelength. This makes them nearly temporally locked and effectively increase the length with strong coupling. Our control pulse happens to be fifth harmonic of the main pulse because of the easy access of 400-nm pulses. 


\begin{figure}[htbp]
\centering
\includegraphics[width=0.45\textwidth]{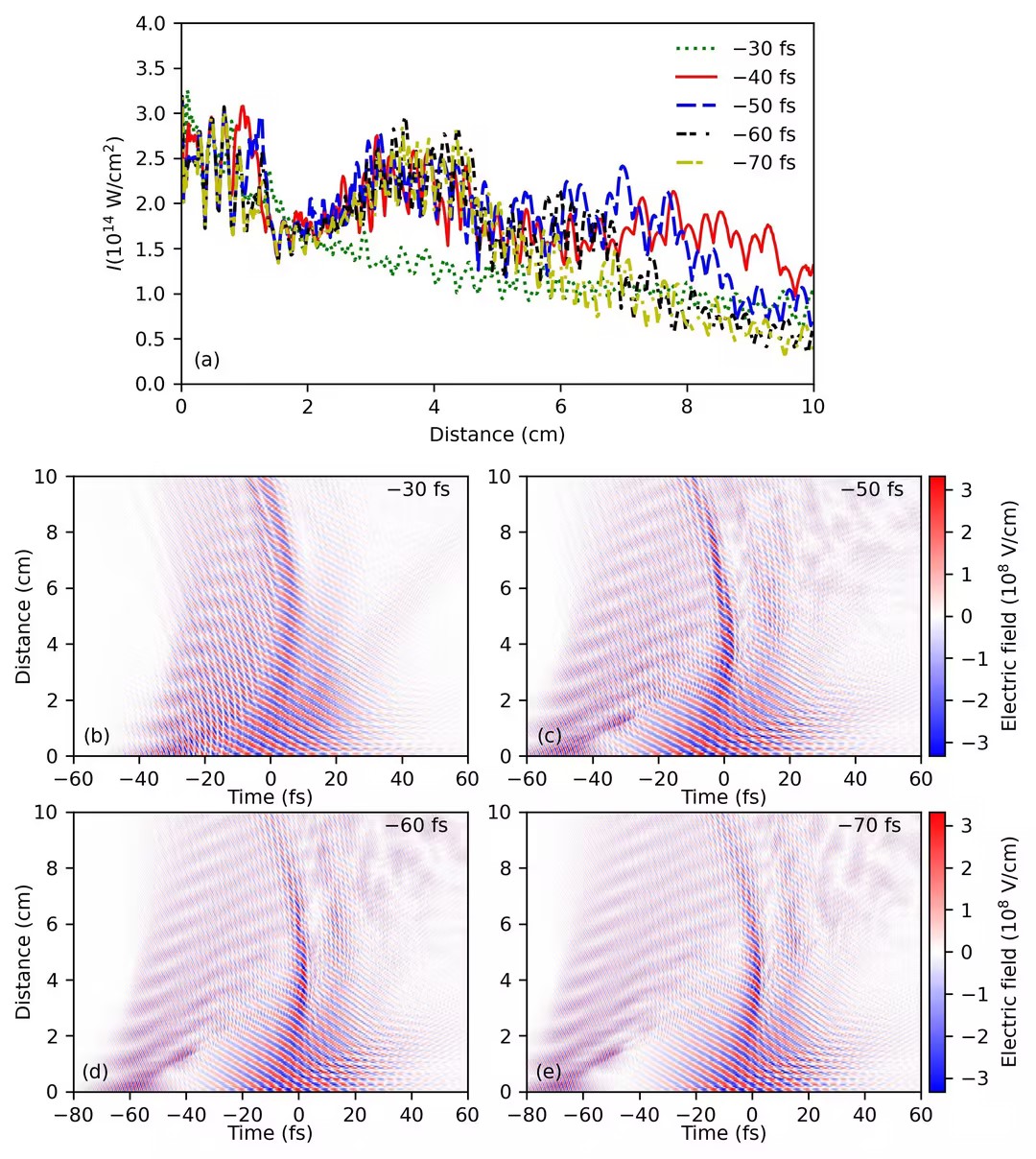}
\caption{Effects of the delay on the propagation dynamics. (a) Peak intensity versus distance for different delay. Spetral evolution over a 10-cm distance for a delay of (b) $-30$ fs, (c) $-50$ fs, (d) $-60$ fs, (e) $-70$ fs.}
\label{fig4}
\end{figure}%
The propagation dynamics for time delays between $-70$\,fs and $-30$\,fs with a 10\,fs increment are shown in Fig.~\ref{fig4}. This initial delay, which determines the timing that the main pulse and the control pulse are temporally overlapped, is critical for an efficient coupling. A delay of $-40$\,fs gives best extended length. The spectral evolution for a delay of $-30$ fs, $-50$ fs, $-60$ fs, and $-70$ fs is shown in Fig.~\ref{fig4}(b)-(e), respectively. At the optimal delay of $-40$\,fs, the control pulse interacts with the main pulse when the main pulse is near the maximum intensity of self-compression. The high intensity and short pulse duration enables a strong cross-phase modulation than that at a delay of $-30$\,fs. A longer negative time delay obviously decouples the two pulses. At a time delay of $-70$\,fs, the evolution of the peak intensity is quite similar to that without the control pulse.  

\begin{figure}[htbp]
\centering
\includegraphics[width=0.45\textwidth]{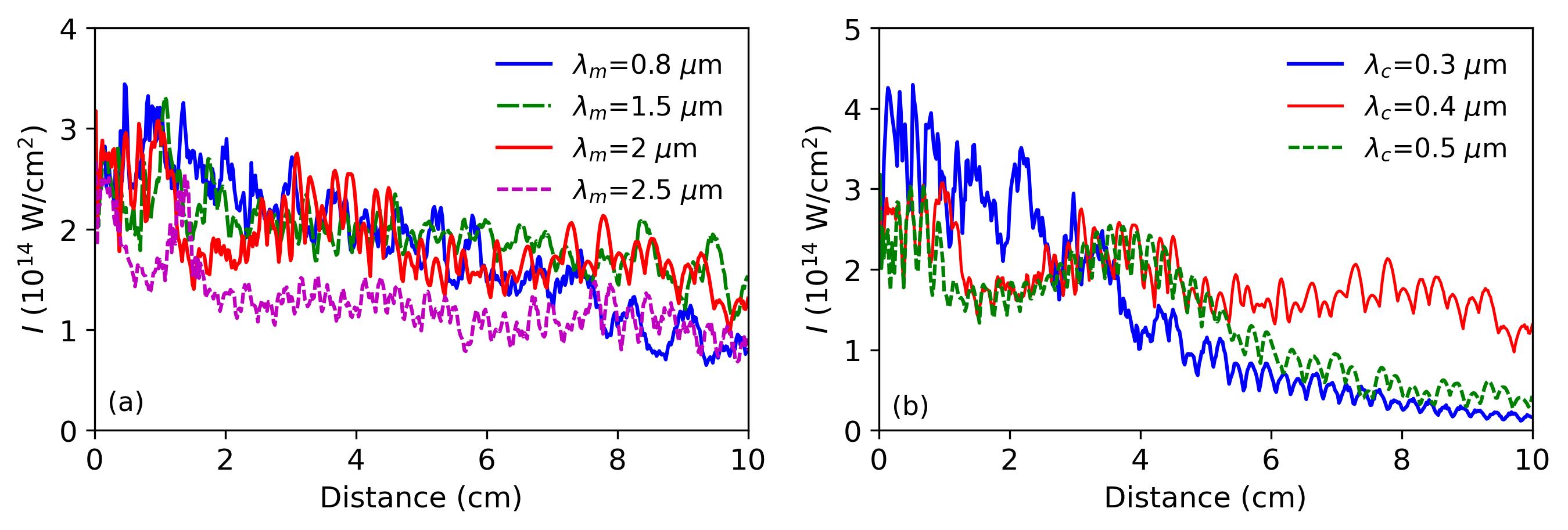}
\caption {Peak intensity versus propagation distance for (a) different main pulse wavelengths, (b) different control pulse wavelengths.}
\label{fig5}
\end{figure}%

Figure~\ref{fig5}(a) show the evolution of the peak intensity with different driving laser wavelengths when the control pulse remains at 400-nm. With an 800-nm driving pulse, the intensity is overall decreasing. Self-compression is only observed for the case of 2-$\mu$m. Anomalous dispersion is required for soliton self-compression~\cite{SchadeOE2021}. At shorter driving wavelengths, the fundamental mode is not in that region. At longer wavelengths, the fundamental and higher-order capillary modes have a greater group velocity mismatch. The temporal walk-off brings down the intensity and is harmful for the self-compression~\cite{Nagar2021OE}. The effect of the control pulse wavelength is shown in Fig.~\ref{fig5}(b), where the wavelength of the main pulse is maintained at 2-$\mu$m. The peak intensity is barely affected by a 500-nm control pulse. Using a 300-nm control wavelength, The large difference in the group velocity of the main and control pulses makes the coupling inefficient. Strong modulation occurs at the first few centimeters probably due to self-focusing. After that the intensity decays quickly. 

\begin{figure}[htbp]
\centering
\includegraphics[width=0.45\textwidth]{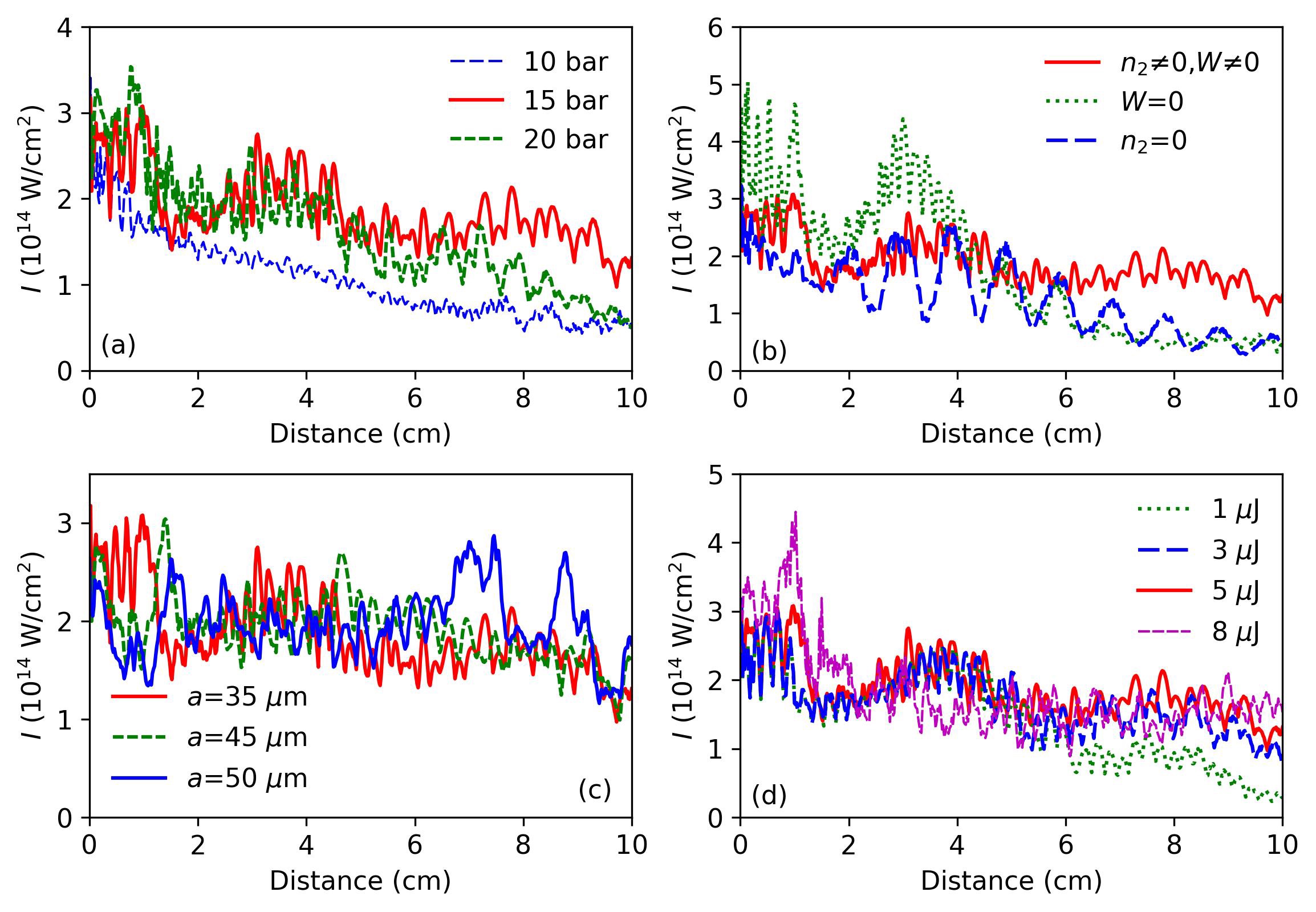}
\caption {Parameter dependencies of the peak intensity evolution versus propagation distance: (a) gas pressure, (b) role of ionization and nonlinear refractive index $n_2$, (c) capillary size, (d) energy of the control pulse. }
\label{fig6}
\end{figure}%
Other parameter dependencies are presented in Fig.~\ref{fig6}. 
Figure~\ref{fig6}(a) displays the variation of peak power with propagation distance under different pressures. The stable propagation requires a delicate balance between nonlinearity and dispersion. As the pressure decreases, the nonlinear length decreases while the dispersion length changes little, so the soliton self-compression is absent at $10$\,bar. A higher pressure of 20 bar does not increase the intensity. It is probably because the higher plasma density decreases the intensity. Figure~\ref{fig6}(b) shows that both the ionization and Kerr nonlinearity are indispensable for this stable propagation. The intensity decreases when either the ionization or nonlinear refractive index is switched off. Figure~\ref{fig6}(c) demonstrates the impact of the capillary radius. With different capillary radius, the modal dispersion and modal loss change and modify the propagation dynamics. At a larger core radius, the group velocity mismatch between fundamental and higher-order capillary modes is smaller. In addition, the loss coefficient is also smaller. This allows efficient coupling for higher-order modes. Figure~\ref{fig6}(d) shows the dependency on the control pulse energy.  An energy of 1-$\mu$J is too low to maintain the intensity, while an energy of 3-$\mu$J remains effective to regulate the intensity.  

\section{Conclusion}
We demonstrate that a strong infrared pulse can propagate with high light intensity for an extended length in gas-filled capillaries with the assist of a weak control pulses. The delay is the most crucial parameter in this process. At a suitable delay, efficient self-compression and cross-phase modulation occur simultaneously, propagating the high laser intensities over a longer distance. The effective interaction length is almost doubled from 5-cm to 10-cm, and it is ultimately limited by the modal loss of the capillaries.
The flexibility of the two-pulse method allows it to be applied to diverse scenarios, providing tailored enhancements at specific distance or global optimization. These results may find applications in enhancing the supercontinuum spectrum of infrared driving pulses, the self-compression of energetic pulses, and the generation of high-order harmonics.

\begin{acknowledgments}
This work was supported by Natural Science Foundation of Zhejiang Province (LY19A040005).
\end{acknowledgments}


%

\end{document}